# Study Of Internal Response Of Epoxy Due To Compressive Load Via Experiment And Simulation Using Abaqus FEA Software.

Irfan Dwi Aditya[1, a], Widayani[2,b], Sparisoma Viridi[3,c], Siti Nurul Khotimah[4,d]

[1-4] Nuclear and Biophysics Research Division, Institut Teknologi Bandung, Jalan Ganesha No.10, Bandung, West Java, Indonesia

[a]irfan.aditya@s.itb.ac.id, [b]widayani@fi.itb.ac.id, [c]dudung@fi.itb.ac.id, [d]nurul@fi.itb.ac.id

**Abstract.** Epoxy is widely used primarily as a matrix material in the manufacture of polymer matrix composite. Epoxy behavior under compression load has to be understood before the mechanical behavior of polymer matrix composite can be accurately predicted. Simulation model combined with experiment and image analysis have been used to investigate internal response of epoxy polymeric materials subjected to compressive loads. In the experiment, small carbon-based material rods are inserted in the epoxy. The samples are held in one side and subjected to compressive load on the other side. All the samples swell at load sides. Image analysis on the carbon-based rods figures out the internal response, which seems to be isotropy in lateral direction. The results are compared to simulation results. The simulation is conducted using Abaqus FEA software. Similar condition is obtained when a brittle thin material is stuck to the top of simulation model.

## Introduction

Polymeric materials play an important role in modern industries today. Due to their lightweight and mechanical characteristics, polymer materials are being used in many applications especially in mechanical engineering. Epoxy is one type of polymeric material that is widely used primarily as a matrix material in the manufacture of composites. Unlike metals, where conducting different types of loading tests and specimens geometries do not present a significant problem due to homogeneity and isotropy of metals, in polymeric material and polymer matrix composite, this type of testing is challenging because of the inhomogeneities and anisotropy[1].

Another aspect in which the relationship between load and deformation in epoxy resins is more complex than in metals is that the hydrostatic component of the stress on epoxy resins has a significant effect on material response. Furthermore, epoxy behavior under compression load has to be understood before the mechanical behavior of polymer matrix composite can be accurately predicted. Although there is considerable literature on the behavior of epoxy with regard compression[1-2], most of them present macroscopical responses but rarely discuss about internal responses. Plastic deformation followed by barreling in cubic specimens has been observed in other studies[3-5]. In this paper, simulation model combined with experiment and image analysis are used to investigate internal response of epoxy resins polymeric materials subjected to compressive loads. The simulation is carried out using Abaqus FEA software which is based on finite element method. To know the internal response of epoxy due to compressive load, pencil rods, which are carbon-based rods, are used as indicators. The rods are chosen because they are black in color so that they can be clearly distinguished from the epoxy. The rods also easy to fracture and totally straight in shape, so the response of the samples during compressive testing can be considered as the response of epoxy solely and the internal shape changes can be seen through the change of the rods.

## Experimental

Epoxy resin used in this experiment is manufactured by Bratachem, Indonesia. Specimens were made with dimensions of 2 cm × 2 cm × 5 cm. While each carbon-based rods has a diameter of 2 mm and a length of 5 cm. A serie of samples were made with different filler numbers: 1, 2, 3, 5, and 9. Compression test is performed using UCT-5T instrument (illustrated in Fig. 1), which is commonly used for mechanical tests.

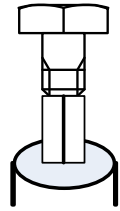

**Fig 1. Compression test of the specimen**

The testing was carried out in ambient temperature ($23^oC$) and 50% RH humidity. The compression was performed using extension rate of 0.5 mm/min. Micro-CT scan images were taken for the sample after compressive testing.

**Simulation**

The dimensional geometry of the simulation model is similar to the geometry of specimens tested in the experiment. The mechanical data inputted in the simulation are obtained from experiment. It was assumed that the carbon-based rods are stick perfectly to the epoxy. The simulations are performed by extend the model along the z-axis up to 1 cm, as with in the experiment. Visualisation of the simulation is given in Fig. 2.

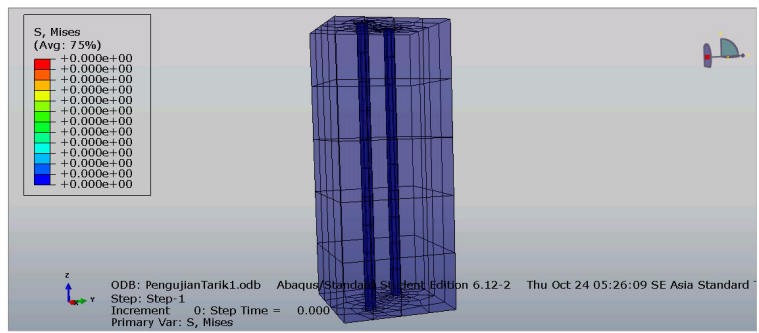

**Fig 2. Simulation model on Abaqus FEA**

**Results**

**Experiment.** As can be seen in the Fig. 3, each specimen shows a common change, ie swell at the top (load side). This figure also clearly shows that the carbon-based rods are bent, which indicates that the expansion is happened in entirely epoxy samples.

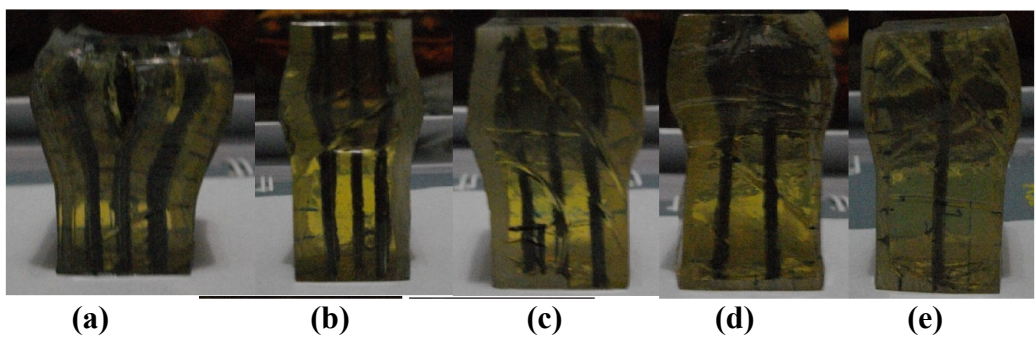

(a)    (b)    (c)    (d)    (e)

**Fig 3. The samples after compressive testing, for different amount of filler:**
**[(a) 9, (b) 5, (c) 3, (d) 2, and (e) 1]**

Young Modulus and ultimate stress obtained from experiment are presented below.

**Table 1. Young Modulus and Ultimate Stress**

| Total Filler in a sample | Young Modulus (MPa) | Ultimate Stress (MPa) |
|---|---|---|
| 1 | 1613,8 | 66,439 |
| 2 | 1580,8 | 69,451 |
| 3 | 1542,0 | 69,020 |
| 5 | 1559,6 | 66,261 |
| 9 | 1640,4 | 64,607 |
| Average | 1587,32 | 67,1556 |

Table 1 shows that there is no significant differences both in Young Modulus and Ultimate Stress of all the samples. As each sample has different total filler number, this means that total filler number does not affect Young Modulus and Ultimate Stress as expected because the filler (pencil rods) are very easy to fracture, so the response of the samples during compressive testing can be considered as the response of epoxy only.

To investigate internal picture of the samples, some samples were scanned using Micro computer tomography instrument. The results show that all the filler are broken at many locations. For example, the picture obtained for a sample containing 3 fillers is shown by Fig. 4.

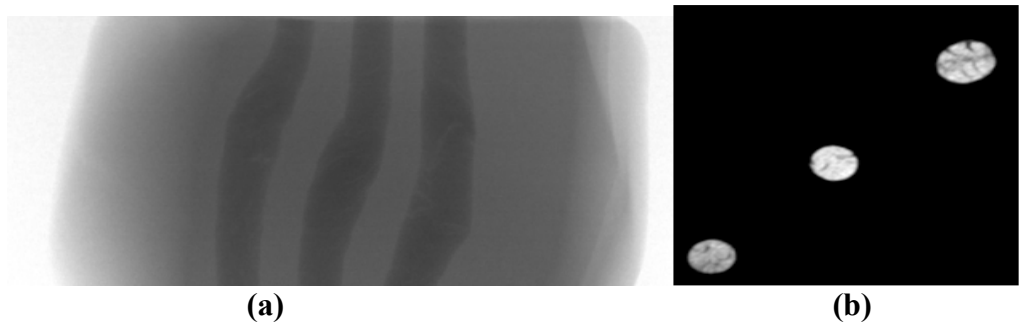

(a)          (b)

**Fig 4. Micro CT scan result (image pixel size : 78.38 μm): (a) A side view of the sample that have cracks, (b) Top cross-sectional view of the sample that have cracks**

It can be seen that in fact, the filler rods are broken at many locations, which means that actually during the compression testing the response of the samples can be considered only from the epoxy.

**Simulation.** We perform the simulation using Abaqus FEA software by conducting the same procedure as in the experiment, that is an isostrain procedure, and the results are as follow.

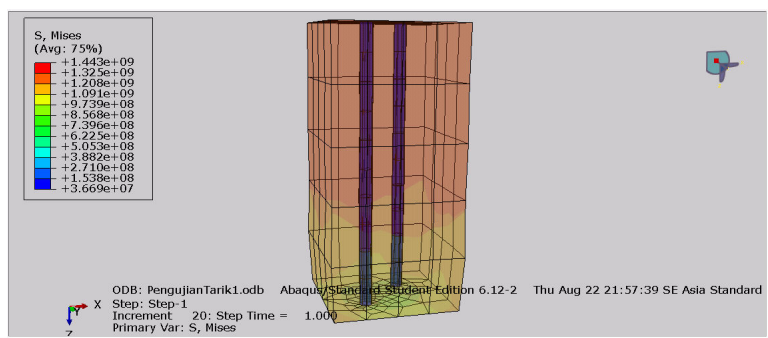

**Fig 5. Result obtained from simulation using Abaqus FEA**

If we perform normal isostrain procedure in the simulation, the result obtained from Abaqus FEA shows no expansion at the top region of the model as seen in Fig. 5, thus it is inconsistent with the experimental results. This inconsistency appears maybe because in the experiment the sample is stuck to the pressure object from the compressing tools.

To improve the simulation result, we add a brittle thin material to cover the upper part of the model. This thin material is stuck to the model. Using this procedure, the simulation result for a sample containing 2 fillers is shown by Fig. 6 below.

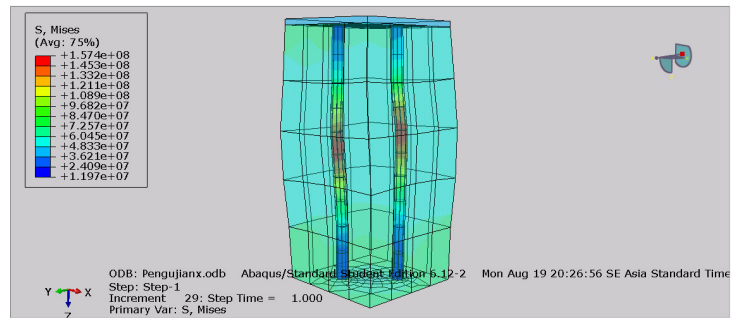

Fig 6.Verification of the model from the experiment

As we can see in the Fig. 6, the simulation result gives similar response to experimental results, i.e. the model swells as with experimental results.

**Summary**

Internal response of the epoxy due to compressive load has been studied via experiments (compressive testing and image characterization using CT-Scan instrument) and simulation using Abaqus FEA software. Under compressive load, epoxy swells at location near to compressive load acts. Image analysis using CT-Scan Instrument shows that carbon-based rods fillers are broken at many locations, which indicates that during the compression testing the response from the samples can be considered as from the epoxy only. The image also shows that the swell seems to be isotropy in lateral direction. Similar result is obtained from simulation when a brittle thin material is stuck to the top of the simulation model.

**Acknowledgement**

Authors would like to extend their thanks to Riset Inovasi dan KK (RIK) ITB with contract no. 122.64/AL-J/DIPA/PN/SPK/2013 for supporting this research.